\documentclass[%
 aip,
 jcp,
 amsmath,
 amssymb,
reprint,
]{revtex4-1}

\usepackage{graphicx}
\usepackage[dvipsnames]{xcolor}
\usepackage{dcolumn}
\usepackage{bm}
\usepackage[utf8]{inputenc}
\usepackage[T1]{fontenc}
\usepackage{mathptmx}

\begin{document}

\title{Excitation and characterization of long-lived hydrogenic Rydberg states of nitric oxide}

\author{A. Deller}
\altaffiliation[Present address: ]{Max-Planck-Institut f\"ur Plasmaphysik,\\
Boltzmannstra\ss e 2, 85748 Garching bei M\"unchen, Germany}
\author{S. D. Hogan}
\email{s.hogan@ucl.ac.uk}
\affiliation{%
 Department of Physics and Astronomy, University College London, \\ Gower Street, London, WC1E 6BT, United Kingdom
}

\date{\today}

\begin{abstract}
High Rydberg states of nitric oxide (NO) with principal quantum numbers between $40$ and~100 and lifetimes in excess of $10~\mu$s have been prepared by resonance enhanced two-color two-photon laser excitation from the X\,$^2\Pi_{1/2}$ ground state through the A\,$^2\Sigma^+$ intermediate state. Molecules in these long-lived Rydberg states were detected and characterized $126~\mu$s after laser photoexcitation by state-selective pulsed electric field ionization. The laser excitation and electric field ionization data were combined to construct two-dimensional spectral maps. These maps were used to identify the rotational states of the NO$^+$ ion core to which the observed series of long-lived hydrogenic Rydberg states converge. The results presented pave the way for Rydberg-Stark deceleration and electrostatic trapping experiments with NO, which are expected to shed further light on the decay dynamics of these long-lived excited states, and are of interest for studies of ion-molecule reactions at low temperatures. 
\end{abstract}

\pacs{}

\maketitle 

\section{Introduction}\label{introduction}

Highly excited Rydberg states of small diatomic molecules, for example, NO, N$_2$ and O$_2$, play important roles in ionization, recombination and collision processes in plasma environments and planetary atmospheres.~\cite{giusti_multichannel_1980, golubkov_chemical_2011, golubkov_microwave_2012, beigman_collision_1995} In the laboratory, high-resolution laser and microwave spectroscopy of Rydberg states allow the most accurate determination of molecular ionization and dissociation energies.~\cite{holsch19a} In zero kinetic energy (ZEKE) photoelectron spectroscopy~\cite{reiser88a} the long lifetimes and susceptibility to ionization of Rydberg states of neutral molecules in weak electric fields, are exploited to accurately determine the rotational-, vibrational- and fine-structure of cations. This circumvents space-charge limitations which arise in experiments performed directly with the ions. The concept of using a Rydberg electron to shield an ion from space charge and stray electric fields can also be extended to low-energy, or low-temperature, studies of ion-molecule reactions. This was first implemented in the moving frame of reference of a single pulsed supersonic beam.~\cite{pratt94a} More recently, significant advances have been made in the study of the $\mathrm{H}_2^+ + \mathrm{H}_2 \rightarrow \mathrm{H}_3^+ + \mathrm{H}$ reaction at collision energies corresponding to $E_{\mathrm{kin}}/k_{\mathrm{B}}\simeq300$~mK, by exploiting the methods of Rydberg-Stark deceleration to merge molecules traveling in two separate beams. ~\cite{allmendinger16a}

Rydberg-Stark deceleration involves the use of inhomogeneous electric fields to exert forces on atoms or molecules in Rydberg states with large static electric dipole moments.~\cite{hogan_rydberg-stark_2016} In merged beam collision studies, chip-based Rydberg-Stark decelerators~\cite{hogan12b,allmendinger14a} are used to transport samples of Rydberg molecules  in one beam, onto the axis of a second beam of target molecules, with a precisely controlled relative velocity. In these devices the molecules are confined in continuously moving electric traps. If these traps are decelerated and brought to rest in the laboratory-fixed frame of reference the resulting samples can be used for studies of slow excited-state decay processes that occur on timescales between $10~\mu$s and 1~ms.~\cite{selier11a,hogan13a,seiler16a,zhelyazkova19a} Rydberg-Stark deceleration and trapping experiments have been performed with H, D, He and H$_2$.~\cite{hogan08a,hogan09a,hogan13a,allmendinger13a,lancuba_electrostatic_2016,zhelyazkova19a} However, it is of interest to extend these studies to heavier and more complex molecules. To achieve this it is necessary to identify appropriate laser photoexcitation schemes to prepare Rydberg states with large static electric dipole moments, and lifetimes greater than $\sim10~\mu$s. Here we demonstrate that long-lived Rydberg states with large electric dipole moments can be prepared in NO by two-color two-photon excitation from the X\,$^2\Pi_{1/2}$ level. By combining laser photoexcitation spectra and electric-field ionization data recorded $126~\mu$s after photoexcitation, the observed series of long-lived Rydberg states have been characterized. Information was obtained on their principal quantum numbers, $n$, and the rotational states of the NO$^+$ ion core to which they converged. 

High Rydberg states of NO have been the subject of numerous experimental and theoretical investigations (see, for example,~Refs.\citenum{ono_higher_1980,seaver83a,ebata83a,kaufmann_rydberg_1985,reiser88a,pratt_twophoton_1989,vrakking_lifetimes_1995,bixon96a,vrakking96a, mccormack98a,goodgame_stark_2002,patel_observation_2007,patel_rotational-state-selective_2007,jones08a}). The work reported here builds on these detailed studies of the structure and dynamics of Rydberg states of NO, many of which were carried out in particular in the context of the development of ZEKE photoelectron spectroscopy,~\cite{reiser88a} and more recent studies of plasma formation in cold Rydberg gases.~\cite{morrison08a,morrison_very_2009,morrison_dynamics_2015} The $\mathrm{NO}^+$ cation is stable and has a closed-shell structure. Consequently, for a Rydberg electron with sufficiently high orbital angular momentum, i.e., for $\ell\gtrsim3$ ($\ell$ is the orbital angular momentum quantum number), the Rydberg states in NO have similar characteristics to those of a one electron atom. However, the charge distribution associated with a Rydberg electron in a state with $\ell\lesssim3$ can destabilize the bond in the NO$^+$ cation, resulting in dissociation. Because of the high density of states of different rotational and vibrational series close to the adiabatic ionization threshold, the decay dynamics of any individual Rydberg state depends strongly on its energetic position and effects of external fields.~\cite{chupka_factors_1993, clarson_stark_2008} The detailed characterization of intramolecular interactions between these excited states is therefore essential in preparing samples suitable for deceleration and trapping. 

In the following, the intramolecular interactions in high Rydberg states of NO are discussed in Section~\ref{sec:RydbergNO}. In Section~\ref{sec:expt} the apparatus and methods used in the experiments are described. Experimental studies of long-lived Rydberg states of NO, with values of $n$ between~40 and~100 are presented in Section~\ref{sec:results}. Finally, in Section~\ref{sec:conc} opportunities to exploit the long-lived Rydberg states studied here for Rydberg-Stark deceleration and trapping experiments are discussed before conclusions are drawn.


\section{High Rydberg states of NO}\label{sec:RydbergNO}

In the absence of external fields the energies, $E_{n, \ell, N^+, v^+}$, of the high Rydberg states of NO can be expressed to first order in the Hund's case (d) angular momentum coupling scheme as
\begin{equation}\label{eqn:rydberg}
\frac{E_{n, \ell, N^+, v^+}}{h c} = W_{v^{+}} + B^+_{v^{+}}\, N^{+}(N^{+} + 1) - \frac{R_{\mathrm{NO}}}{\left[n - \delta_{\ell(N^+,\,v^+)}\right]^2},
\end{equation}
where $W_{v^{+}}$ is the ionization wavenumber of the Rydberg series converging to the ground electronic state of the NO$^+$ cation with vibrational quantum number $v^+$, $B^+_{v^{+}}$ and $N^+$ are the rotational constant and rotational quantum number of the ion core, the Rydberg constant for  NO is $R_{\mathrm{NO}} = 109735.31$~cm$^{-1}$, and $\delta_{\ell(N^+,\,v^+)}$ represents the Hund's case (d) quantum defect. For the $v^+=0$ states of interest here $B^+_0 = 1.9842$~cm$^{-1}$,~\cite{vrakking96a} and the case (d) quantum defect, which in general depends on the value of $v^+$,~\cite{bixon96a,goodgame_stark_2002} is denoted $\delta_{\ell(N^+)}$.

The dissociation rate of a Rydberg NO molecule depends strongly on the Rydberg electron charge density close to the ion core. For any particular series of Rydberg states this charge density scales with $n^{-3}$. However, for each value of $n$ it also depends strongly on the value of $\ell$. For excited states with $\ell \gtrsim 3$, the centrifugal barrier that separates the Rydberg electron from the core suppresses the interactions between the two.~\cite{vrakking_lifetimes_1995,bixon96a} High-$\ell$ states are therefore much less susceptible to dissociation and exhibit sharp Rydberg series with characteristics similar to those of a one-electron atom.

The selection rules for electric-dipole transitions restrict the maximum value of $\ell$ of the Rydberg states directly accessible by laser photoexcitation from the ground state of NO. For the 
\begin{eqnarray}
n\ell N^+ \;\left[\mathrm{X}^+\,^1\Sigma^+\right]\leftarrow\mathrm{A}\,^2\Sigma^+\leftarrow\mathrm{X}\,^2\Pi_{1/2}
\end{eqnarray}
resonance-enhanced two-color two-photon excitation scheme used here,~\cite{} the states excited must have $\ell<4$. However, weak applied, or stray, electric fields can mix long-lived high-$\ell$ character into these otherwise short-lived low-$\ell$ states. This $\ell$-mixing can allow access to hydrogenic Rydberg states in the photoexcitation process. Rapidly switching off such fields after excitation has been shown to `lock' the excited molecules into these states.~\cite{held_lifetime_1998} High-$|m_{\ell}|$ and hence high-$\ell$ character can also be acquired as a result of charge-dipole interactions between the excited Rydberg molecules and stray ions present at the time of laser photoexcitation,~\cite{merkt_lifetimes_1994,vrakking_lifetimes_1995, palm_ion_1997, held_role_1998} or electric dipole-dipole interactions.~\cite{seiler16a} In principle, quantum-state-selective preparation of similar long-lived states, and even circular Rydberg states for which $|m_{\ell}|=\ell=n-1$, could be achieved in molecules like NO using the method of crossed electric and magnetic fields~\cite{delande88a,hare88a,zhelyazkova16a} and its variants,~\cite{morgan18a} or by adiabatic microwave transfer.~\cite{hulet_rydberg_1983, liang_circular_1986,cheng94a} However, these schemes have so far only been implemented in atoms. 

\subsection{Intramolecular interactions}\label{sec:interactions}

The details of the angular momentum character, lifetimes and predissociation dynamics of the high Rydberg states in NO depend strongly on intramolecular interactions between different ionization channels. These interactions are induced by the electric dipole and quadrupole moments, and electric polarizability of the NO$^+$ ion core.~\cite{bixon96a} Here they are briefly summarized to aid in the interpretation of the experimental data presented in Section~\ref{sec:results}. 

{\bf (i-a)} In a coordinate system with its origin at the center of mass of the molecule, the NO$^+$ cation has a static electric dipole moment of $\sim1$~D.~\cite{jungen70a,bergeman74a,gray93a,glendening95a,bixon96a} The interaction of a Rydberg electron with this internal electric dipole moment in the neutral NO molecule results in the mixing of Hund's case (d) states for which~\cite{bixon96b}
\begin{eqnarray}
\Delta\ell = \pm1;~\Delta N^+ = \pm1;~\Delta N = 0; \textrm{ and } \Delta M_N = 0,
\end{eqnarray} 
where $N$ and $M_N$ are the total angular momentum quantum number excluding spin, and the associated azimuthal quantum number of the neutral NO molecule, respectively.~\cite{bixon96a} The strength of this type of intramolecular interaction depends on the inverse square of the distance between the Rydberg electron and the ion core. It is therefore of most significance for low-$\ell$ Rydberg states. In this case states for which $\ell\lesssim3$.

{\bf (i-b)} The higher order electric multipole moments of the NO$^+$ ion core lead to further mixing of the Hund's case (d) basis states. The effect of these additional interactions between the Rydberg-electron and the ion core are that, e.g., in the case of the charge-quadrupole interaction, Rydberg states for which~\cite{bixon96b} 
\begin{eqnarray}
\Delta\ell=0,\pm2;~\Delta N^+=0,\pm2; \textrm{ and } \Delta N = 0,
\end{eqnarray} 
are mixed. As for the interactions denoted type (i-a) above, these type (i-b) couplings are strongest for values of $\ell\lesssim3$. 

{\bf (ii)} The multielectron character of the NO$^+$ ion core leads to a further coupling of the Hund's case (d) basis states. This is the s$\sigma$--d$\sigma$ configuration mixing~\cite{fredin87a} ($\sigma$ represents the $\lambda=0$ projection of $\vec{\ell}$ onto the internuclear axis). The consequence of this interaction is that even in the absence of external fields pure $n$s or $n$d Rydberg eigenstates do not exist in NO. The corresponding states therefore exhibit mixed s and d character. Since this configuration mixing only occurs for states with the same value of $N$, the resulting s-d mixed eigenstates also have $\Delta N^+ = \pm2$ character. s$\sigma$--d$\sigma$ mixing plays a major role in determining the values of the Hund's case (d) quantum defects for the corresponding states. However, because the resulting $\ell$- and $N^+$-mixed Rydberg states cannot be directly excited with the laser photoexcitation scheme used here, these states do not play a direct role in the interpretation of the experimental data. 

\subsection{Effects of external electric fields}\label{sec:fields}

In addition to intramolecular couplings between distinct series of Rydberg states, external electric fields can cause further mixing of the Hund's case (d) basis states in NO. In weak electric fields this mixing does not directly couple Rydberg series converging to different rotational states of the NO$^+$ ion core, but does couple Rydberg states with the same value of $N^+$ which differ in their values of $\ell$ by $\pm1$. The couplings induced by a static electric field between the Hund's case (d) basis states require that~\cite{vrakking96a,patel_observation_2007}
\begin{eqnarray}
\Delta\ell = \pm1;~\Delta N = 0, \pm1;~\Delta N^+ = 0; \textrm{ and }\Delta M_N = 0.
\end{eqnarray}
In experiments of the kind described here weak electric fields, on the order of 10~mV/cm generally arise from incomplete stray field compensation at the position of laser photoexcitation.~\cite{deller18a} Time-varying stray fields with amplitudes between 10 and 100~mV/cm can also result from the presence of ions at the time of photoexcitation.~\cite{merkt_lifetimes_1994} An important effect of electric fields of this scale on the Rydberg states of NO is to impart $n$f character into the highly-degenerate manifold of higher $\ell$ hydrogenic states. This $\ell$-mixing enables direct access to the hydrogenic states with quantum defects $\delta_{\ell(N^+)}=0$ in the laser photoexcitation process. The Hund's case (d) quantum defect of the $n$f(2) states is $\delta_{\,\mathrm{f}(2)}=0.01$.~\cite{bixon96a} For $n=40$ an electric field of 10~mV/cm (100~mV/cm) imparts $\sim0.01$\% ($\sim1$\%) 40f character into the outermost $\ell$-mixed hydrogenic $n=40$ states with negative Stark energy shifts. At $n=60$, fields of 10~mV/cm result in $\sim1$\% 60f character in the corresponding states. Fields of 100~mV/cm transfer $\sim5$\% 60f character to all $n=60$ Stark states with negative Stark shifts. This form of $\ell$-mixing only allows direct photoexcitation of Stark states with negative Stark shifts. However, because the stray fields can be time varying (i.e., they are associated with electric field noise, or the presence of ions that can travel rapidly through the laser excitation region) the excited molecules can evolve non-adiabatically into a range of Stark states with negative and positive Stark shifts. 

\begin{figure}
\includegraphics[width=0.5\textwidth]{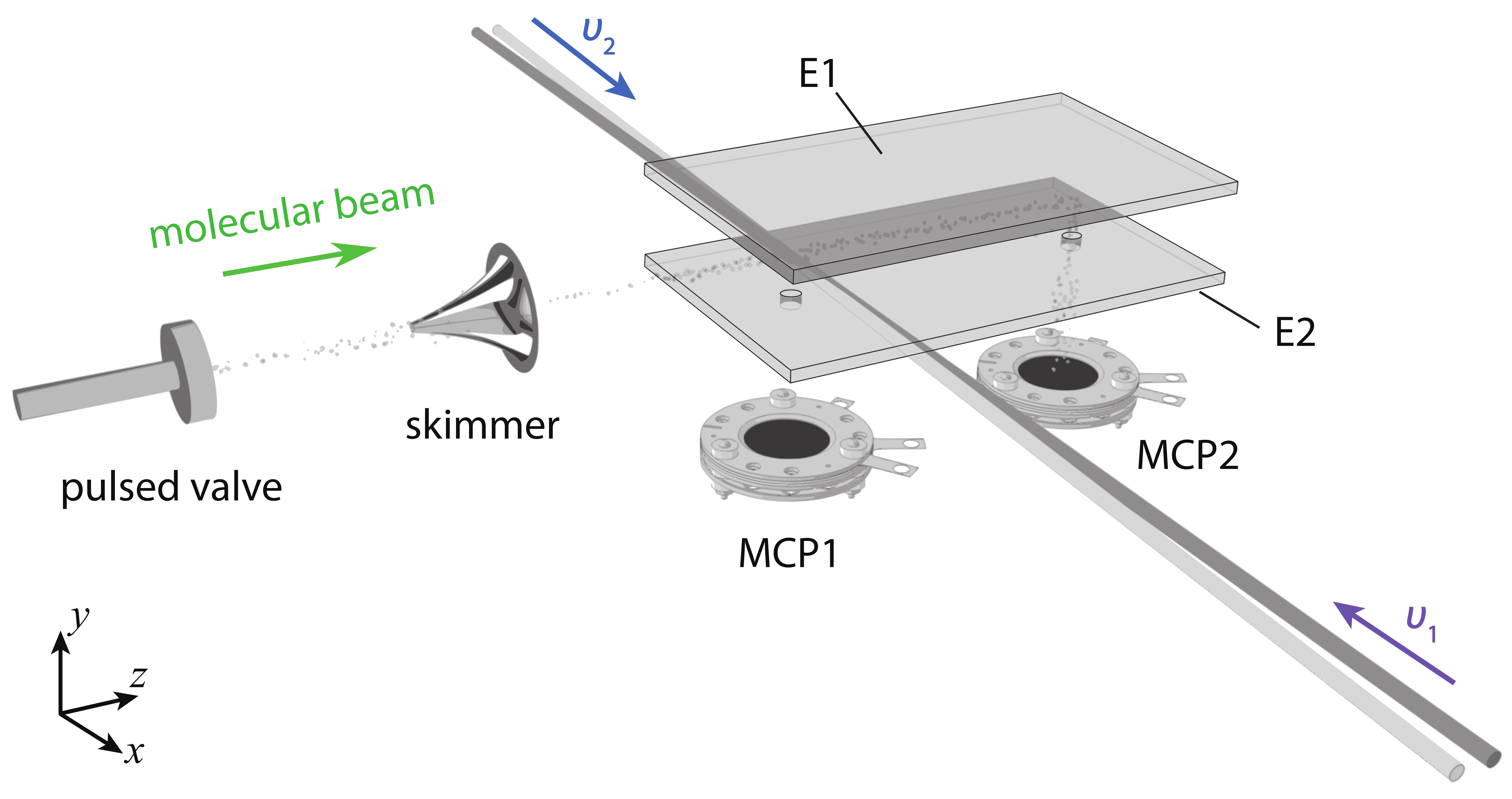}
\caption{\label{fig:schematic} Schematic diagram of the experimental arrangement. NO molecules in a skimmed, pulsed supersonic beam were prepared in high Rydberg states between electrodes E1 and E2 using a resonance-enhanced two-color two-photon excitation scheme with laser radiation at wave numbers $\upsilon_1$ and $\upsilon_2$. The excited molecules could be detected by pulsed electric field ionization immediately after photoexcitation, or after a flight-time of 126~$\mu$s, with the resulting ions or electrons collected on MCP1 or MCP2, respectively.}
\end{figure}

\section{Experiment}\label{sec:expt}
\subsection{Experiment\label{techniques}}

A schematic diagram of the apparatus used in the experiments is presented in Figure~\ref{fig:schematic}. NO molecules were emitted into vacuum from a reservoir maintained at a pressure of 1.4~bar through a pulsed valve operated at a repetition rate of 25 Hz. The resulting supersonic molecular beam with a mean longitudinal speed of 790~m$\,$s$^{-1}$ passed through the 2-mm-diameter aperture of a skimmer and entered a volume vertically bounded by two copper plates, E1 and E2 in Figure~\ref{fig:schematic}. These planar electrodes were 60~mm wide in the $x$ dimension and 150~mm long in the $z$ dimension. They were orientated parallel to each another with a separation in the $y$ dimension of 6.7~mm.

Between the electrodes the molecular beam was intersected by the $\upsilon_1=44\,194$~cm$^{-1}$ (226.27~nm) frequency-tripled output of a Nd:YAG-pumped pulsed dye laser. This laser beam had an energy of $\sim20\,$--$\,80~\mu$J per pulse and was collimated to a waist of 2~mm. In the following, the time at which the molecules interacted with this 8-ns-duration laser pulse is denoted $t_0$. The frequency doubled output of a second dye laser, coincident and counterpropagating with respect to this first beam, also crossed the molecule beam. This had an energy of $\sim1.2$~mJ per pulse, was compressed to a waist of 2~mm, and could be tuned over the range from $\upsilon_2=30\,460$ -- $30\,520$~cm$^{-1}$ (328.30 -- 327.65~nm). The fundamental frequencies of the two dye lasers were monitored using a fibre-coupled wavelength meter with an absolute accuracy of $\pm0.005$~cm$^{-1}$ ($\pm150$~MHz). 

After laser photoexcitation Rydberg NO molecules could be detected upon the application of pulsed ionization potentials to electrode E1. The resulting electrons or ions were then accelerated thorough one of two 5-mm-diameter apertures in E2 and collected on microchannel plate (MCP) detectors. This detection procedure could be implemented at two positions in the electrode structure. Molecules in Rydberg states and photoions could be detected directly in the laser excitation region with the resulting charged particles accelerated through a first aperture in E2 to MCP1. Molecules excited to long-lived Rydberg states, with lifetimes $>10~\mu$s, could be detected by delayed pulsed electric field ionization at a second position between E1 and E2, 100~mm downstream from the first. In this case the electrons were collected on MCP2. 

\begin{figure}
\includegraphics[width=0.4\textwidth]{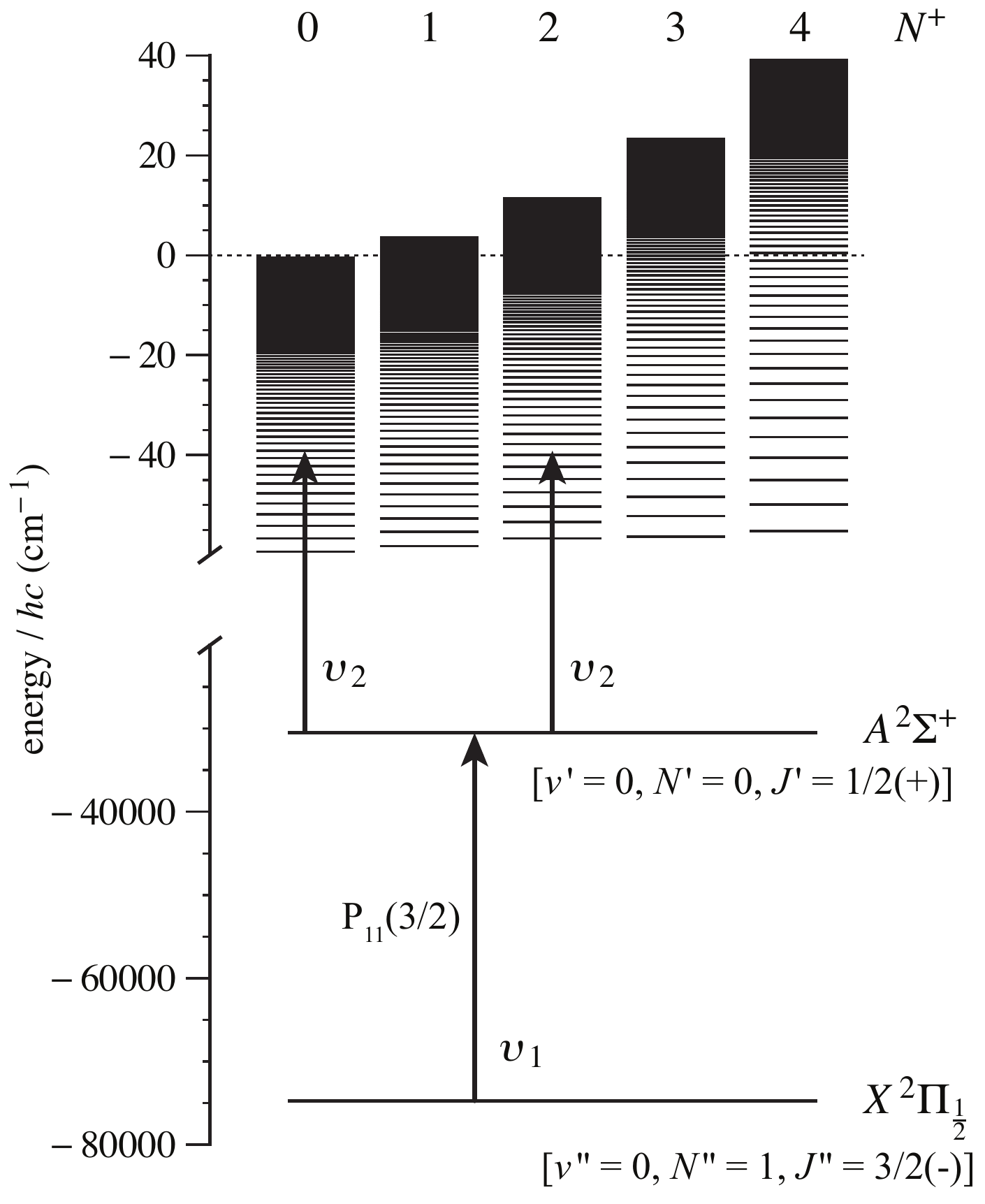}
\caption{\label{fig:NO_levels} Resonance-enhanced two-color two-photon scheme used to excite NO molecules from the X\,$^2\Pi_{1/2}$ ground electronic state to $n$p$(0)$ and $n$f$(2)$ Rydberg states. The scale on the vertical axis indicates the energy with respect to the X$^+$\,$^1\Sigma^+$ state of NO$^+$ with $\nu^{+} = 0$ and $N^{+} = 0$.}
\end{figure}

\subsection{Laser photoexcitation scheme\label{sec:excite}}

\begin{figure}
\includegraphics[width=0.48\textwidth]{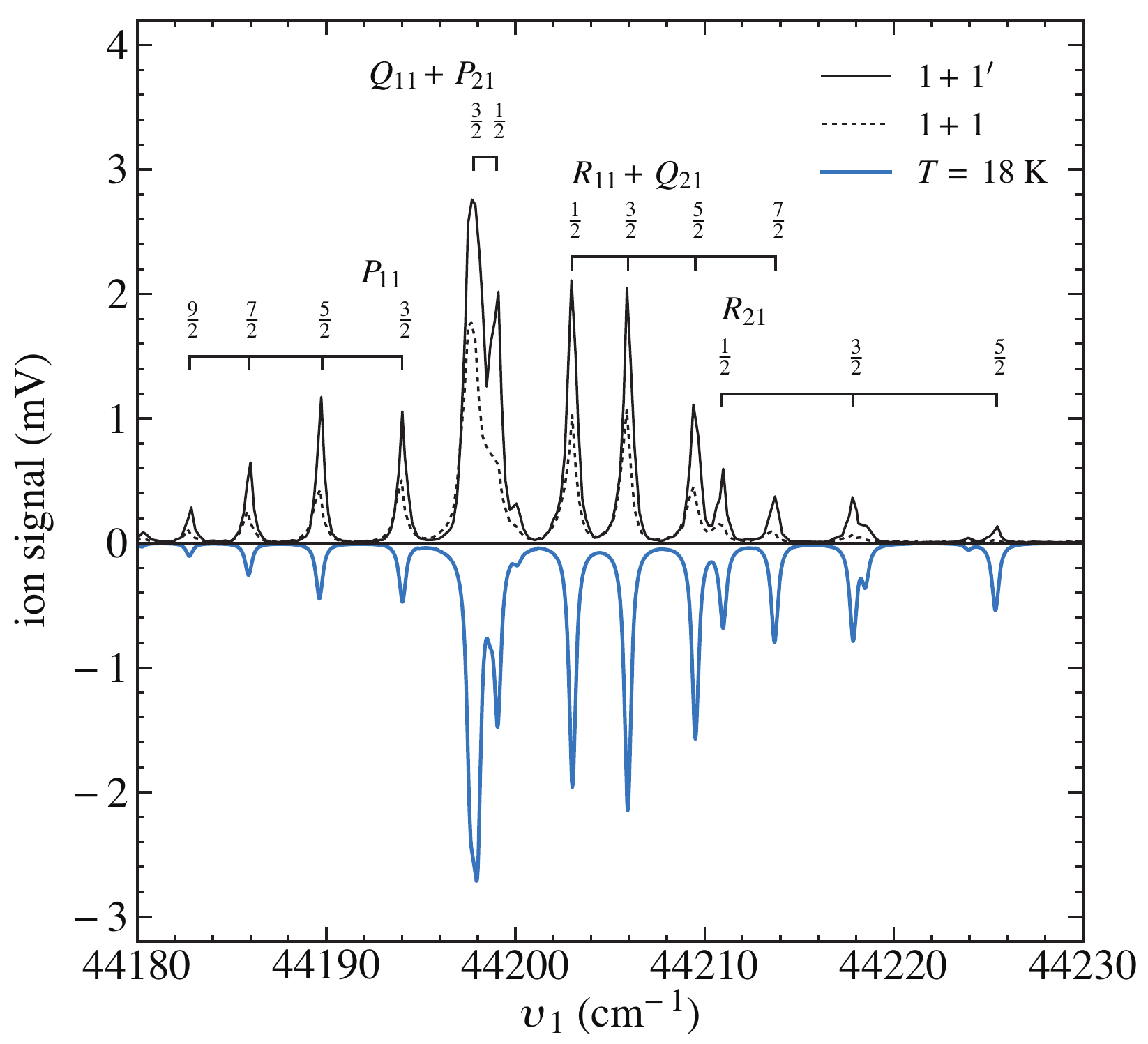}
\caption{\label{fig:A_state} One-color, $1 + 1$, and two-color, $1 + 1^\prime$ REMPI spectra of the $\mathrm{A}\,^{2}\Sigma^{+}  (\nu' = 0) \leftarrow \mathrm{X}\,^{2}\Pi_{\frac{1}{2}} (\nu'' = 0)$ transitions in NO. A calculated spectrum for a rotational temperature of 18~K is displayed as the inverted curve beneath the experimental data.}
\end{figure}

The resonance-enhanced two-color two-photon excitation scheme used here to prepare high Rydberg states of NO has been widely studied in the literature~\cite{ebata83a,seaver83a,reiser88a,pratt_twophoton_1989,mccormack98a,patel_observation_2007,morrison08a} and depicted is in Figure~\ref{fig:NO_levels}. Molecules initially in the $\mathrm{X}\,^{2}\Pi_{1/2}$ $\nu^{\prime \prime} = 0$, $N^{\prime \prime} = 1$, $J^{\prime \prime} = \frac{3}{2}(-)$ level were excited via the $P_{11}(\frac{3}{2})$ transition to the intermediate $A\, ^{2}\Sigma^{+}$ $\nu^{\prime} = 0$, $N^{\prime} = 0$, $J^{\prime} = \frac{1}{2}(+)$ state. This transition occurred at $\upsilon_1 = 44\,193.988(10)$~cm$^{-1}$. To characterize the molecular beam, and this step of the laser photoexcitation scheme, one-color ($1 + 1$) resonance enhanced multi-photon ionization  (REMPI) spectra of the $A ^{2}\Sigma^{+}  (\nu' = 0) \leftarrow \mathrm{X}\,^{2}\Pi_{\frac{1}{2}} (\nu'' = 0)$ transitions were recorded with a laser pulse energy of $\sim80~\mu$J. The results are displayed in Figure~\ref{fig:A_state} (dashed grey curve). The features in this spectrum were assigned as indicated in the figure following from Ref.~\citenum{hippler_detection_1995}. 

Further measurements were made using a two-color ($1 + 1^\prime$) REMPI scheme with a second laser of wavenumber $\upsilon_2 = 30\,534.94$~cm$^{-1}$ tuned 12.496~cm$^{-1}$ above the adiabatic ionization threshold. For these measurements the $\upsilon_2$ pulse energy was reduced to $\sim20~\mu$J to minimise the $1 + 1$ REMPI process. The spectral resolution in this two-color ($1 + 1^\prime$) scheme (continuous black curve in Figure~\ref{fig:A_state}) is higher than in the one-color ($1 + 1$) case because this lower laser pulse energy used to drive the A\,$^{2}\Sigma^{+} \leftarrow \mathrm{X}\,^{2}\Pi$ transition did not cause saturation. A spectrum calculated~\cite{engleman__1971, western_pgopher:_2017} for a sample of molecules with a rotational temperature of 18~K (inverted blue curve) yields good quantitative agreement with the experimental data. The $P_{11}(\frac{3}{2})$ transition measured to lie at $\upsilon_1 = 44\,193.988(10)$~cm$^{-1}$ is clearly identified as an isolated and well-resolved resonance. For all subsequent experiments reported here, $\upsilon_1$ was set to this transition wavenumber.

The second laser used to drive transitions from the A$\,^{2}\Sigma^{+}$ state to high Rydberg states could be tuned over the range from $\upsilon_2 = 30\,460$ to $30\,520$~cm$^{-1}$ to access $n$p(0) and $n$f(2) states [$n\ell(N^+)$] with values of $n$ between 40 and 100. In general states with $n$p character dissociate rapidly (i.e., in $< 1$~ns\:~\cite{vrakking_lifetimes_1995}). Consequently, the longer-lived $n$f(2) states are expected to dominate the experimental spectra. 


\subsection{Electric field ionization\label{sec:ion}}

In the experiments, pulsed electric fields with a rise time of $\sim 30$~ns were applied to accelerate photoions produced by REMPI to MCP1 (see Figure~\ref{fig:schematic}). Similar fields were used to ionize neutral Rydberg molecules for detection immediately after photoexcitation. To detect molecules in long-lived Rydberg states, slowly-rising pulsed ionization fields~\cite{ducas75a,patel_rotational-state-selective_2007} (rise time $\sim2~\mu$s) were applied $126~\mu$s after laser photoexcitation. In these measurements the electric field amplitude at the time of ionization was correlated with the electron signal detected at MCP2 to provide information on the internal quantum state populations.

In Rydberg states of a non-hydrogenic atom or a molecule electric field ionization occurs in two ways.~\cite{gallagher_rydberg_1994} The first, adiabatic electric field ionization, is encountered for states with low values of $\ell$, or more precisely $\ell$-mixed Rydberg-Stark states with low values of $|m_{\ell}|$, i.e., $|m_{\ell}|\lesssim3$. The large zero-field quantum defects of the low-$\ell$ states that contribute to these low-$|m_{\ell}|$ Stark states result in large avoided crossings in fields at and beyond the Inglis-Teller electric field, $F_{\mathrm{IT}}=F_0/(3n^5)$, where states that differ in their value of $n$ by $+1$ first cross [$F_0=2\,h\,c\,R_{M}/(e\,a_M)$, with $R_M$ and $a_M$ the Rydberg constant and the Bohr radius corrected for the reduced mass]. For the typical rates at which ionization electric fields are switched in the experiments reported here, i.e., $\sim0.2$~V/cm/ns, these large avoided crossings are traversed adiabatically. In fields between $F_{\mathrm{IT}}$ and the ionization electric field, $F_{\mathrm{ion}}$, these states do not undergo significant further Stark energy shifts and the molecules are, on average, not particularly strongly polarized. In these cases, ionization occurs when the saddle point in the potential that arises as a result of the presence of the electric field, crosses the energy of each Stark state. The resulting `classical' ionization field is therefore $F_{\mathrm{ion}}^{\mathrm{adiabatic}}=F_0/(16n^4)$,~\cite{littman78a} which is approximately equal to the field in which the tunnel ionization rate is 10$^8$~s$^{-1}$.~\cite{PhysRevA.99.013421} 

When $|m_{\ell}|\gtrsim3$, the absence of low-$\ell$ components of the Stark states results in small, or almost exact, crossings at and beyond $F_{\mathrm{IT}}$. In this situation, for typical electric field switching rates these crossing are traversed diabatically and once polarized the Stark states maintain their static electric dipole moments and approximately linear Stark energy shifts until ionization occurs. However, the orientation of the static electric dipole moment of the polarized molecules strongly affects their ionization electric field.~\cite{damburg79a,damburg83a} States with negative Stark shifts have dipole moments oriented parallel to the electric field, with the Rydberg electron localized between the ion core and the Stark saddle point. As a result, tunnel ionization occurs when the energy of the saddle point approaches the energy of the Rydberg state. The ionization electric field of the outermost negatively-shifted Stark state is $F_{\mathrm{ion}}^{\mathrm{diabatic}\,^-}=F_0/(9n^4)$. This expression accounts for the Stark energy shift of the state and corresponds to a tunnel ionization rate of $\sim10^8$~s$^{-1}$. 

High-$|m_{\ell}|$ states with positive Stark shifts also ionize following diabatic traversal of the level crossings at and beyond $F_{\mathrm{IT}}$. However, in these states the dipole moment of the molecule is oriented antiparallel to the electric field. As a result the Rydberg electron is localized on the opposite side of the ion core to the Stark saddle point. This remains the case in all fields in which the Stark energy shifts are linear. However, for high electric fields $n$-mixing leads to the depolarization of the electric dipole and a redistribution of electron charge density into the region between the ion core and the Stark saddle point. Under these conditions ionization can occur, but ionization rates of $10^8$~s$^{-1}$ are only reached in fields twice as large as those for the outermost negatively-shifted Stark states. Therefore $F_{\mathrm{ion}}^{\mathrm{diabatic}\,^+}=2F_0/(9n^4)$. States that lie between the outermost Stark states ionize in fields between $F_{\mathrm{ion}}^{\mathrm{diabatic}\,^-}$ and $F_{\mathrm{ion}}^{\mathrm{diabatic}\,^+}$. In the experiments reported here the 126$~\mu$s flight time of the Rydberg molecules to the detection region above MCP2 acts to filter those in long-lived hydrogenic Stark states with $|m_{\ell}|\gtrsim3$. The molecules in these hydrogenic states undergo diabatic electric field ionization. 

In one-electron atoms, e.g., H, D, or Ps, all Rydberg states undergo diabatic electric field ionization provided the energy splittings associated with the fine- and hyper-fine structure are small. In many-electron atoms and in molecules diabatic ionization typically occurs in states for which $|m_{\ell}|\gtrsim3$. These states, with $\delta_{\ell(N^+)}\simeq0$ for all individual angular momentum components, are referred to as hydrogenic.

\begin{figure}
\includegraphics[width=0.47\textwidth]{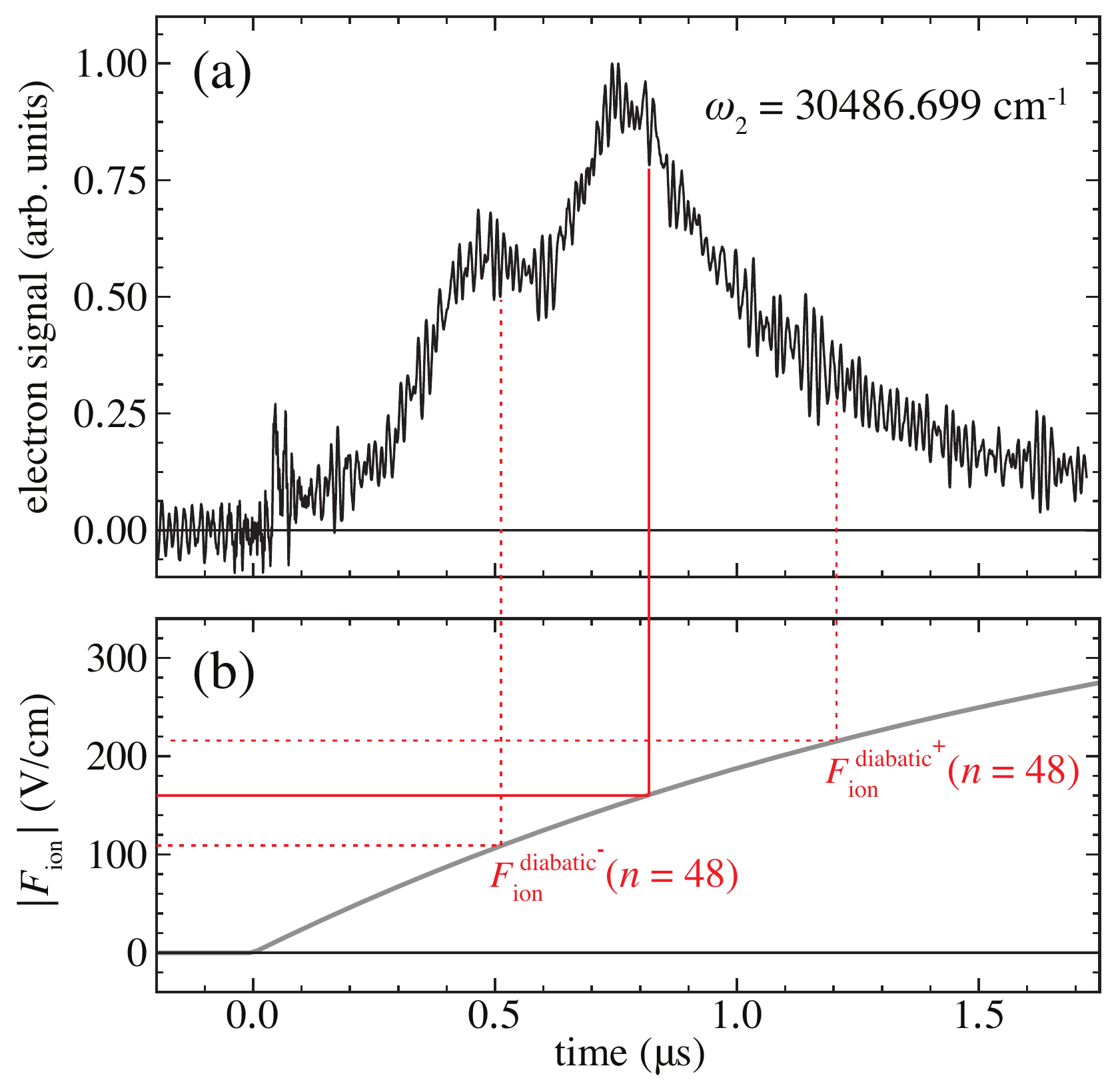}
\caption{\label{fig:mcp_raw} (a) Rydberg state electric field ionization signal recorded upon the application of the slowly-rising ionization field displayed in~(b) $126~\mu$s after laser photoexcitation. When recording the data in~(a) $\upsilon_2$ was set to $30\,486.99$~cm$^{-1}$ to excite predominantly $\ell$-mixed hydrogenic Rydberg states in NO with $n=48$ and $N^+=2$. The red lines connecting panel~(a) and panel~(b) represent the range of ionization electric fields associated with the manifold of $\ell$-mixed hydrogenic Rydberg states with $n=48$ (see text for details).}
\end{figure}

The slowly-rising pulsed potential, $V_\mathrm{E1}(t)$, applied to electrode E1 (see Figure~\ref{fig:schematic}) for detection had the form
\begin{eqnarray}
V_\mathrm{E1}(t) &=& V_\infty  \left[1 - \exp\left(- t / \kappa \right)\right]
\end{eqnarray}
with $V_\infty = -300$~V and $\kappa = 1.84$~$\mu$s. The time dependence of the corresponding ionization field is shown in Figure~\ref{fig:mcp_raw}(b). In this field, which rises to $\sim250$~V/cm in a time of 1.5$~\mu$s, high-$n$ Rydberg states generally ionize at early times and lower-$n$ states at later times. For each value of $n$, the hydrogenic Stark states ionize over a range of times corresponding to ionization fields from $F_{\mathrm{ion}}^{\mathrm{diabatic}^-}$ to $F_{\mathrm{ion}}^{\mathrm{diabatic}^+}$. Ionized electrons are then accelerated to the MCP detector with flight times of 5 -- 30~ns depending upon the particular value of $V(t)$ at the time of ionization.

An example of the field-ionized electron signal recorded following the preparation of NO molecules in hydrogenic Rydberg states with $n=48$ ($\upsilon_2=30\,486.699$~cm$^{-1}$) can be seen in Figure~\ref{fig:mcp_raw}(a). In these data, electrons arrive at the MCP at times between $\sim0.3$ and $\sim1.5$~$\mu$s after activation of the ionization pulse with an intensity maximum at $\sim0.8$~$\mu$s. This indicates that the Rydberg states populated at the time of detection ionize in electric fields ranging from $\sim50$ to $\sim250$~V/cm with the intensity maximum at 160~V/cm. This range of fields corresponds approximately to the ionization fields of the hydrogenic Rydberg states in NO with $n=48$, which are centred upon a field of $\sim160$~V/cm and extend from $F_{\mathrm{ion}}^{\mathrm{diabatic}^-}(n=48)=107$~V/cm to $F_{\mathrm{ion}}^{\mathrm{diabatic}^+}(n=48)=215$~V/cm as indicated by the red lines. The electron signal in fields outside this range, and the additional structure in the data in Figure~\ref{fig:mcp_raw}(a), are indicative of the range of Rydberg states populated at the time of electric field ionization. This distribution of population reflects contributions from intramolecular interactions close to the time of laser photoexcitation, and transitions induced by blackbody radiation in the time between laser photoexcitation and detection by electric field ionization.

In the experimental data presented in the Section~\ref{sec:results}, electron time-of-flight distributions, such as those in Figure~\ref{fig:mcp_raw}, have been transformed and presented as a function of ionization field. Numerical particle trajectory calculations were implemented to determine the electron flight times from the position of ionization to the MCP in the time-varying field.

\begin{figure}
\includegraphics[width=0.47\textwidth]{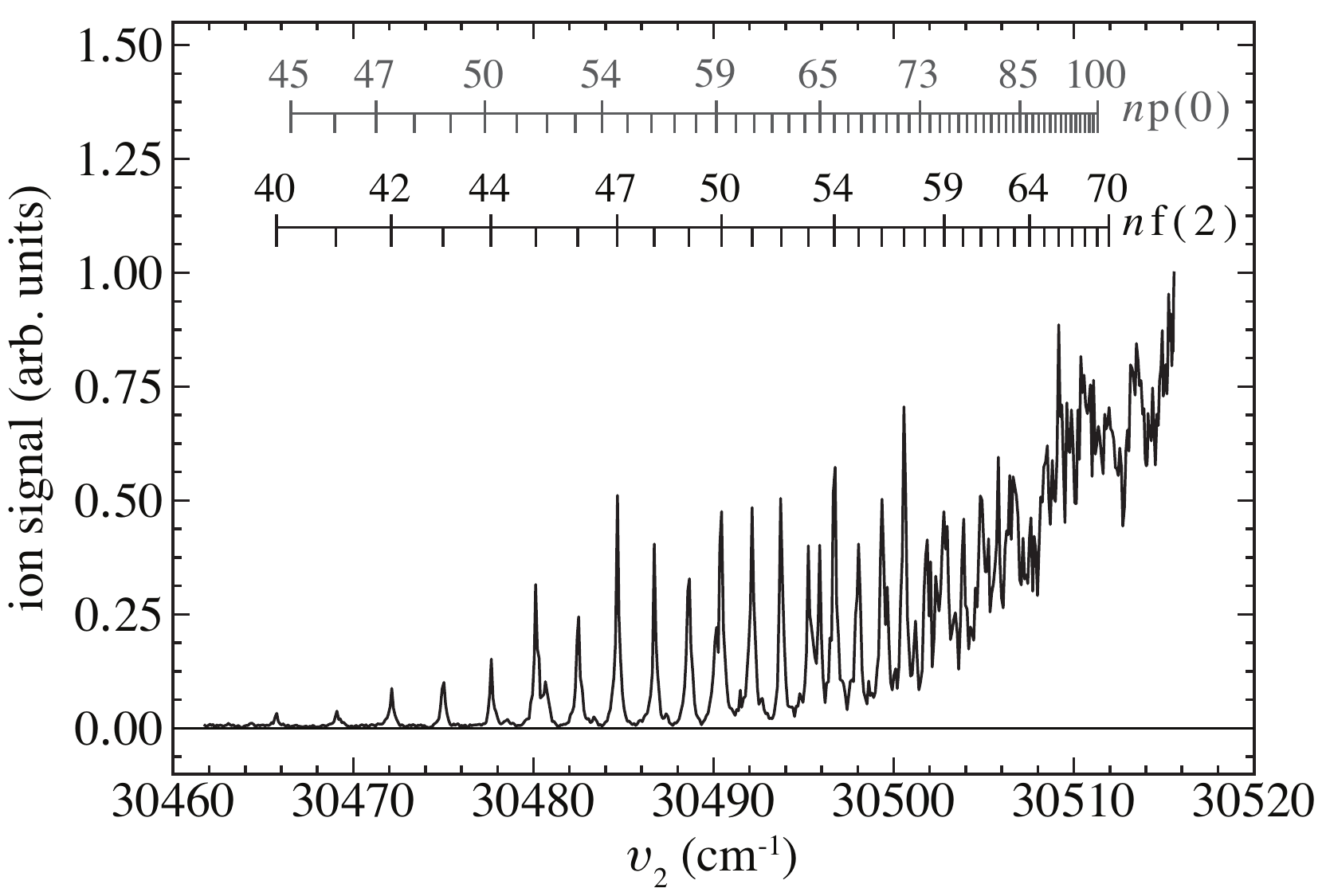}
\caption{\label{fig:wide_n_ions} Total integrated ion signal recorded at MCP1 by pulsed electric field ionization 50~ns after laser photoexcitation. Inset above the data are calculated values of $\upsilon_2$ for transitions to the $n$p(0) and $n$f(2) states.}
\end{figure}

\section{Results}\label{sec:results}

\subsection{Short-lived Rydberg states}

A spectrum recorded following detection of Rydberg NO molecules at MCP1, upon ionization $\sim50$~ns after laser photoexcitation, is displayed in Figure~\ref{fig:wide_n_ions}. Above this spectrum, the wavenumbers associated with the electric dipole allowed transitions to the $n$p(0) and $n$f(2) Rydberg series are indicated. For values of $n$ below 100, the predissociation lifetimes of the $n$p(0) states are $<3$~ns.~\cite{vrakking_lifetimes_1995} Consequently, the features observed correspond predominantly to transitions to $n$f(2) states. Above 30\,500~cm$^{-1}$ transitions to individual Rydberg states are unresolved because of a combination of an increase in the density of states, and effects of $\ell$-mixing in the weak residual uncanceled stray electric fields present at the time of photoexcitation.

\begin{figure}
\includegraphics[width=0.47\textwidth]{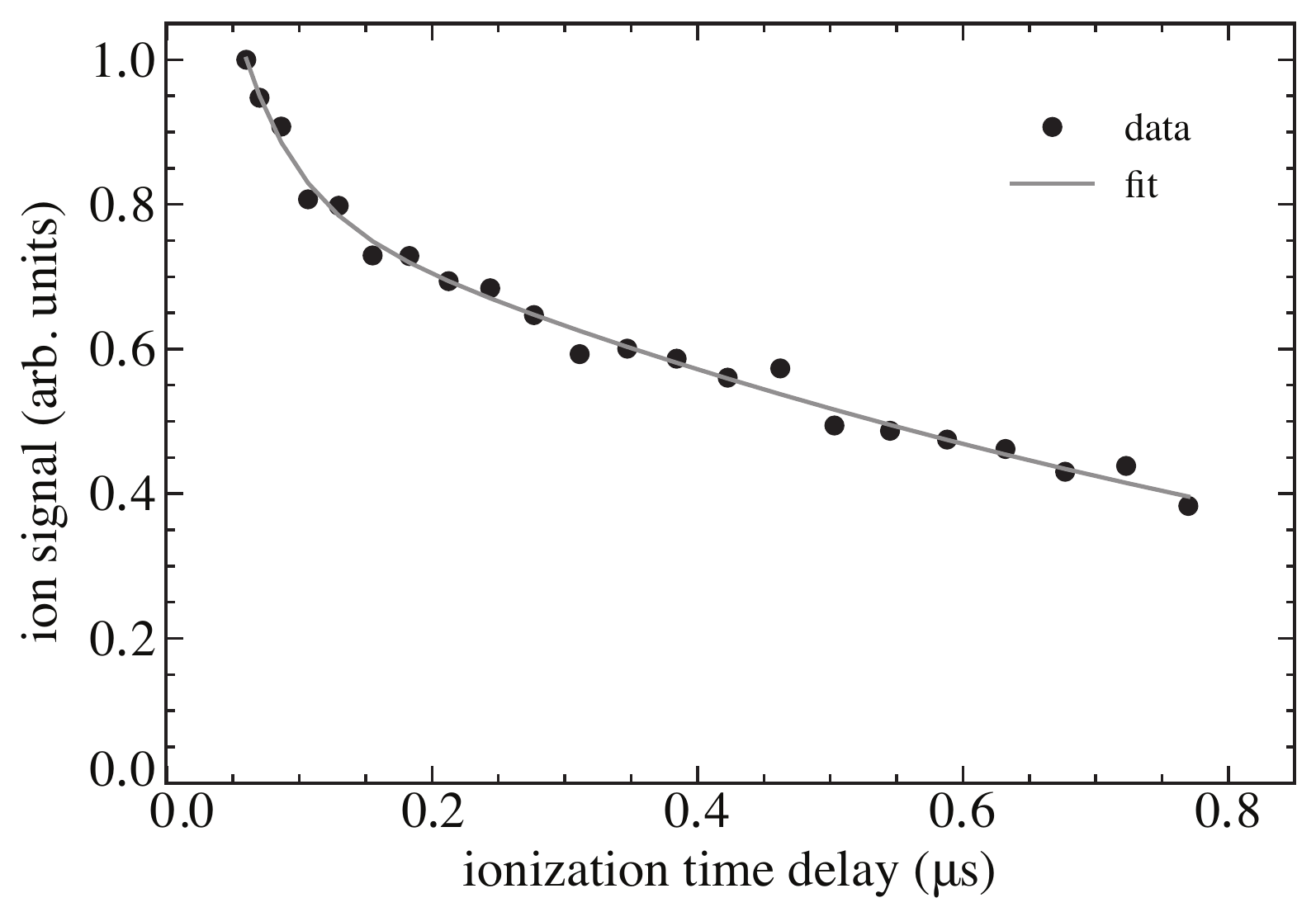}
\caption{\label{fig:lifetime} Total integrated ion signal recorded at MCP1 upon ionization, at a range of times after photoexcitation, of NO Rydberg states with 56f(2) character ($\upsilon_2=30\,499.35$~cm$^{-1}$). The continuous curve represents a two component exponential function fit to the experimental data, with time constants $41\pm10$~ns and $1006\pm50$~ns.}
\end{figure}

The pure $n$f(2) Rydberg states in the spectrum in Figure~\ref{fig:wide_n_ions} have lifetimes between 10 and 100~ns. This can be inferred, for example, from the data in Figure~\ref{fig:lifetime} where for $\upsilon_2= 30\,499.35$~cm$^{-1}$, corresponding to the transition to the 56f(2) state, the total integrated ion signal recorded as a function of time after photoexcitation is displayed. The experimental data in this figure exhibit two dominant decay components. A least-squares fit of a two-component exponential function (continuous curve) yields time constants for these of $41\pm10$~ns and $1006\pm50$~ns. The faster decaying component has a time constant similar to the $\sim25$~ns lifetime of the 56f(2) state reported previously in the literature.~\cite{vrakking_lifetimes_1995} The slow component suggests the presence of molecules in longer-lived states with $\ell\gtrsim3$. Over the 800~ns timescale on which the data in Figure~\ref{fig:lifetime} were recorded, the excited molecules traveled $\sim0.6$~mm. Thus, the dominant contribution to the ionization-time dependent changes observed arises from the decay of the excited molecules and not their motion. 

\begin{figure}
\includegraphics[width=0.47\textwidth]{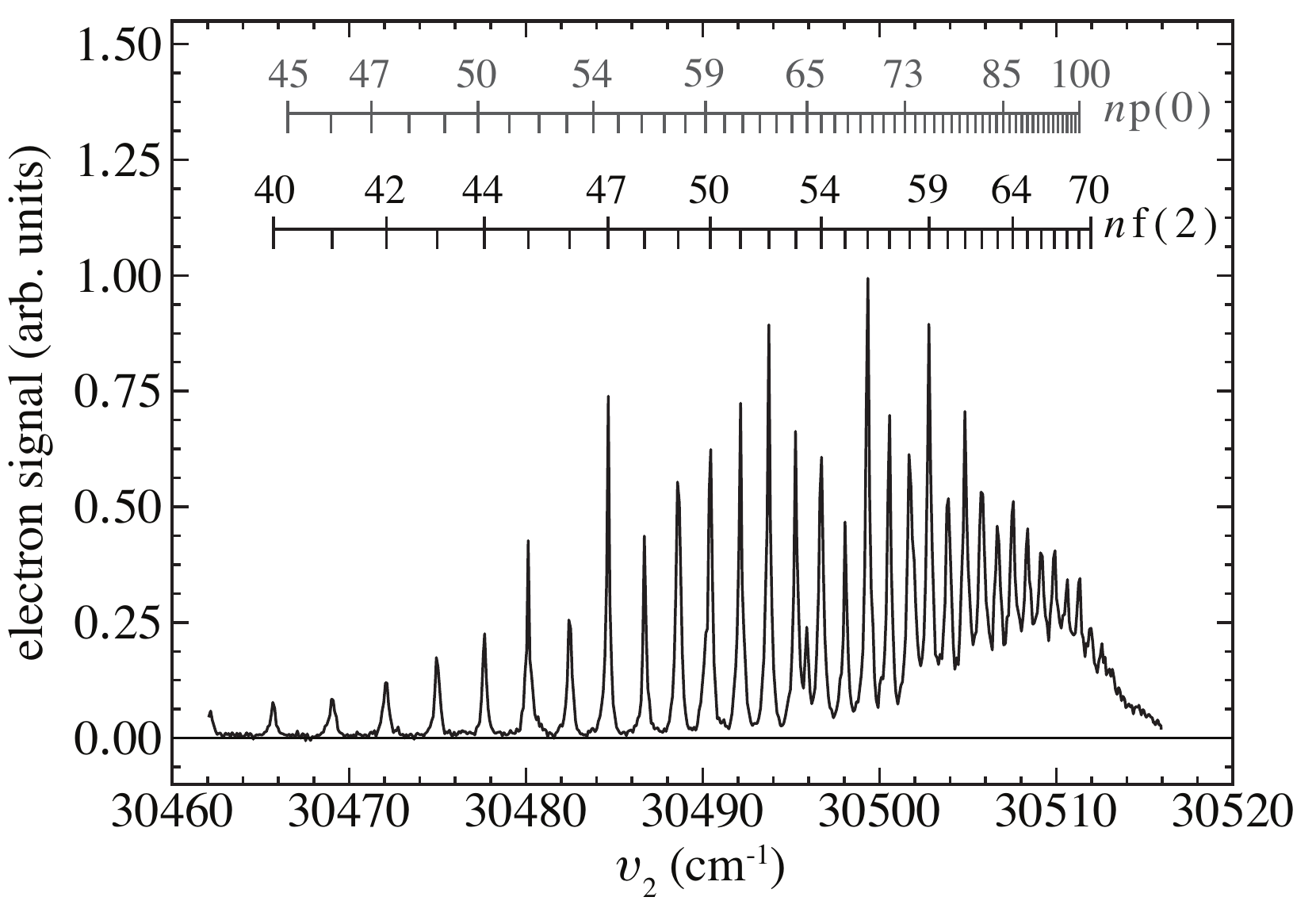}
\caption{\label{fig:wide_n} Total integrated electron signal recorded at MCP2 upon delayed pulsed electric field ionization 126~$\mu$s after laser photoexcitation. Inset above the data are the calculated values of $\upsilon_2$ for the transitions to the $n$p(0) and $n$f(2) states.}
\end{figure}

\subsection{Long-lived Rydberg states}

The spectrum in Figure~\ref{fig:wide_n} was recorded by delayed electric field ionization in the region above MCP2. This spectrum is similar to that in Figure~\ref{fig:wide_n_ions} in that it is dominated by transitions to $n$f(2) Rydberg states converging toward $\upsilon_2 = 30\,534.349(16)$~cm$^{-1}$.  As in Figure~\ref{fig:wide_n_ions} there are features, notably the distinct resonance at 30\,496~cm$^{-1}$, which do not form part of the $n$f(2) series. This resonance lies between the transitions to the 53f(2) and 54f(2) states, and is located at the calculated wavenumber for the transition to the 65p(0) state (see inset). At transition wavenumbers beyond 30\,510~cm$^{-1}$, the observed signal intensity reduces toward zero. This is expected to be a consequence of the higher density of states of different Rydberg series, and the associated increase in the effects of intramolecular interactions and stray electric fields on the lifetimes of the high-$|m_{\ell}|$ hydrogenic Stark states, close to the adiabatic ionization threshold. Lifetime measurements performed with molecules confined in cryogenically cooled electrostatic traps~\cite{hogan09a,lancuba_electrostatic_2016,seiler16a,zhelyazkova19a} are expected to provide further insight into the excited state dynamics in this spectral region. In the absence of external fields, the pure $n$f(2) states~\cite{vrakking_lifetimes_1995} are not sufficiently long lived for an appreciable fraction to survive the 126~$\mu$s flight time to the detection region above MCP2. For detection with reasonable efficiency after this long flight time, the lifetimes of the excited states populated must be $\gtrsim10~\mu$s and consequently the states must have values of $|m_{\ell}|\gtrsim3$. 

\begin{figure*}
\includegraphics[width=0.7\textwidth]{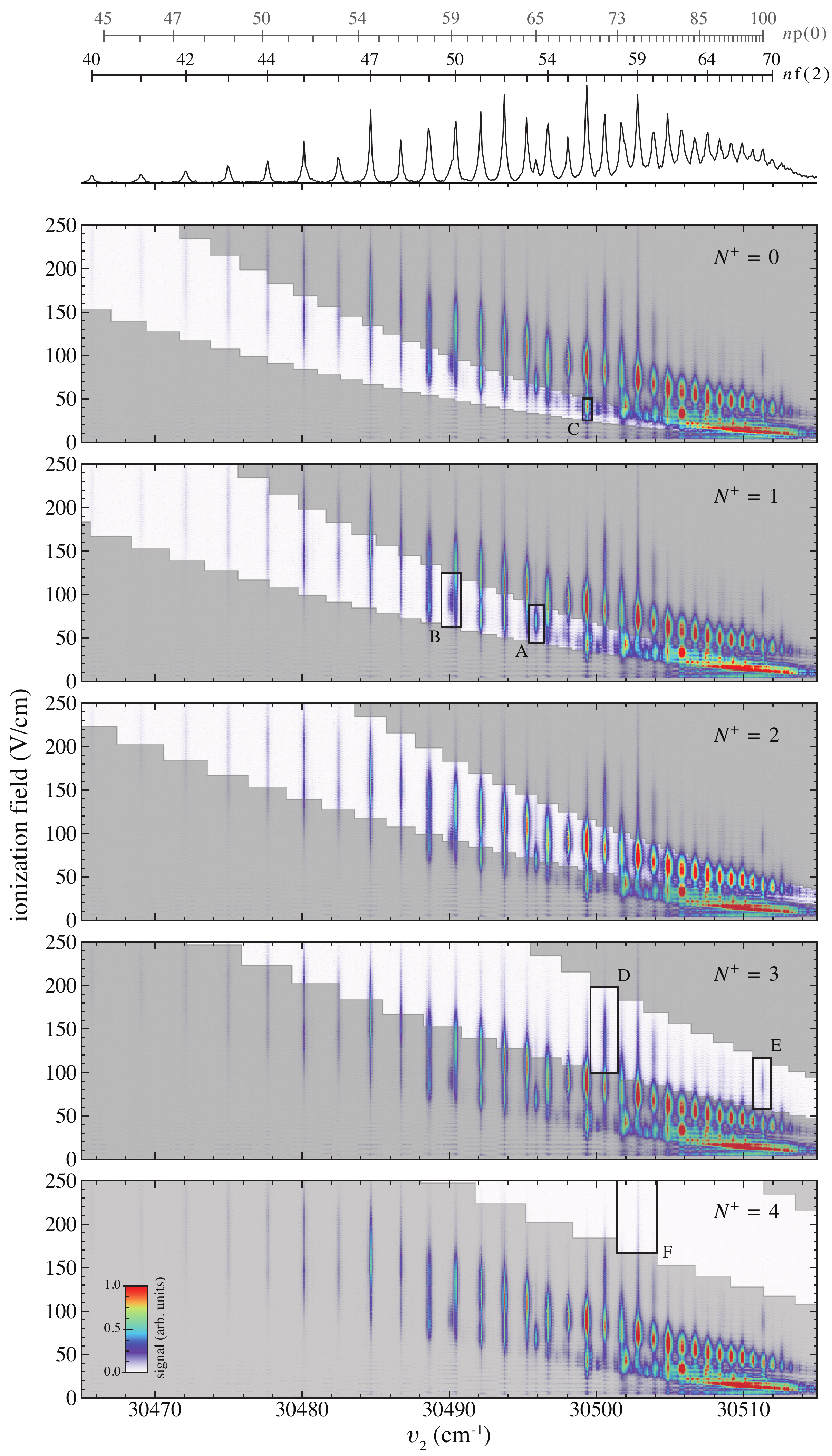}
\caption{\label{fig:wide_n_full} Delayed electric field ionization signal of long-lived Rydberg NO molecules excited in nominally zero electric field. A spectrum representing the total integrated electron signal at each laser wavenumber is displayed in the top panel along with the expected values of $\upsilon_2$ for the transitions to the $n$p(0) and $n$f(2) Rydberg series. The white regions in each of the lower panels correspond to the calculated electric field ionization bands for the hydrogenic Rydberg states of the $N^+=0$, 1, 2, 3 and 4 series (see text for details). The limits of the color scale were chosen to maximise visibility. 
}
\end{figure*}

Additional information on the population of the long-lived excited states in the spectrum in Figure~\ref{fig:wide_n} was obtained by monitoring the electric field ionization signal (as in Figure~\ref{fig:mcp_raw}) at each laser wavenumber across the spectrum. The results of this type of measurement are displayed in Figure~\ref{fig:wide_n_full}. In this figure the total integrated delayed electric field ionization signal (as in Figure~\ref{fig:wide_n}) is included in the top panel. In the two-dimensional spectral maps, the color scale represents the background-subtracted electron signal recorded at each photoexcitation wavenumber. The vertical axis in these panels indicates the ionization electric field to which the observed electron signal can be referred. The experimental data are identical in each of the five panels of the figure. The white bands demarcate the ionization electric fields, ranging from $F_{\mathrm{ion}}^{\mathrm{diabatic}^-}$ to $F_{\mathrm{ion}}^{\mathrm{diabatic}^+}$, of the hydrogenic Rydberg-Stark states with $N^{+} = 0$, 1, 2, 3 and~4. The steps in these white bands are centred at the zero-field photoexcitation wavenumbers of these states.

Comparing the experimental data in Figure~\ref{fig:wide_n_full} to the field ionization bands with which they are overlaid indicates that the most prominent features are consistent with diabatic electric field ionization of hydrogenic Rydberg states with $N^{+} = 2$ (middle panel). Almost all of the resonances observed in the experimental data align with the expected wavenumbers for transitions to this Rydberg series, and the electric field ionization signal associated with these resonances follows the band of ionization fields expected for these hydrogenic states. Because the excited states are long-lived, they must possess negligible $n$s, $n$p or $n$d character, as this would result in rapid pre-dissociation. Therefore they must have values of $|m_{\ell}|\gtrsim3$. Since $\delta_{\ell\geq3(N^+)}<0.01$, the avoided crossings in the Stark maps are sufficiently small that diabatic ionization dominates. Consequently, under the conditions in which the experiments were performed, the ionization electric fields of the long-lived states are expected to reflect those associated with diabatic, rather than adiabatic, ionization. 

The short lifetimes of the pure $n$f(2) Rydberg states~\cite{vrakking_lifetimes_1995}, and the observation that the excited states populated at the time of detection ionize across the full range of electric fields associated with each set of hydrogenic Rydberg-Stark states lead to the conclusion that the long-lived states detected are populated upon mixing with the $n$f(2) states in weak residual uncanceled static or time-dependent electric fields close to the time of photoexcitation as discussed previously in the context of ZEKE photoelectron spectroscopy~\cite{merkt_electric_1993,merkt_lifetimes_1994,palm_ion_1997,held_role_1998}. Since $\delta_{\mathrm{f}(2)}=0.01$, the $n$f(2) states are almost degenerate with the manifold of hydrogenic states with the same value of $n$. For $n=40$ the wavenumber interval between the 40f(2) state and the hydrogenic states of the same series is 0.05~cm$^{-1}$, and below the experimental resolution of 0.17~cm$^{-1}$. Thus for the range of values of $n$ encompassed in the spectrum in Figure~\ref{fig:wide_n_full} the $n$f(2) states and the corresponding hydrogenic states are not resolved, and it is inferred that the $n$f(2) states mix with these longer-lived states even in very weak stray electric fields. This mechanism allows the long-lived hydrogenic states to be partially populated at the time of laser photoexcitation. Only the molecules in these states were then detected by delayed electric field ionization. 

The $n$p(0) states have large quantum defects of $\delta_{\mathrm{p}(0)} \simeq 0.7$.~\cite{vrakking96a} They therefore do not mix so readily with the hydrogenic states with the same value of $n$ and do not enable photoexcitation of long-lived states in the $N^+=0$ series. The short dissociation lifetimes of the $n$p(0) states also mean that Rydberg-Stark states with sufficient $n$p(0)-character to permit laser photoexcitation are, for values of $n\lesssim70$, not sufficiently long-lived to allow detection 126~$\mu$s after excitation. However, it can be seen from the the data in the spectral region between 30\,500 and 30\,515~cm$^{-1}$ in Figure~\ref{fig:wide_n_full}, and the calculated ionization fields of the hydrogenic Rydberg states with $N^+<2$, that for values of $n\gtrsim70$ long-lived states can be populated.

The other features in the experimental data in Figure~\ref{fig:wide_n_full} that do not follow such explicit $n$-dependent behavior can all be assigned to the population of hydrogenic Rydberg states through channel interactions between Rydberg series. Unlike the features attributed to the $N^{+} = 2$ states, which change smoothly with $n$, evidence for the population of Stark states with $N^{+} = 0$, 1, 3 or~4 is more sporadic. For example, the notable interloper in the vicinity of the transition to the 65p(0) state ionizes in the manner expected of the hydrogenic states with $n=60$ in the $N^{+} = 1$ series. This is seen clearly from the overlap of the experimental data with the rectangular box labelled A in the figure. The range of ionization fields associated with this feature are not correlated with any states in the $N^+=0$, 2, 3 or~4 series. A similar feature can be seen as the shoulder on the low wavenumber side of the prominent transition close to the 59p(0) state in the spectrum in the top panel of Figure~\ref{fig:wide_n_full}. Molecules excited at this resonance ionize in the range of fields expected for hydrogenic states with $n=55$ in the $N^{+} = 1$ series (rectangular box labelled B).  The connection between these examples is that the long-lived states of the $N^+=1$ series that are detected are almost degenerate with $n$p(0) states. Thus considering the couplings arising from the intramolecular interaction mechanisms discussed in Section~\ref{sec:interactions}, it is determined that these are examples of the type (i-a) interaction of the Rydberg electron with the static electric dipole moment of the NO$^+$ cation.~\cite{bixon96a} Additional $\ell$-mixing, required to populate the long-lived hydrogenic states of the $N^+=1$ series, must then be induced by weak residual uncanceled electric fields.

In the $N^{+} = 3$ ionization band several features appear in the wavenumber range between  30\,500 and 30\,511~cm$^{-1}$. The strongest of these, indicated by the rectangular boxes labelled D and E, ionize in the fields associated with the hydrogenic states with $n=49$ and $n=56$, respectively. In these cases, the observed states of the $N^+=3$ series are almost degenerate with states in the $n$f(2) series to which electric dipole transitions from the A\,$^2\Sigma^+$ state are allowed in zero electric field. The process by which these long-lived Rydberg states with $N^+=3$ are populated therefore also relies on the channel interaction mechanism of type (i-a). Over the wavenumber range in Figure~\ref{fig:wide_n_full}, one signature of ionization from a state in the $N^+=4$ series is seen. This is a consequence of the near degeneracy of the 59f(2) state with the hydrogenic states with $n=43$ in the $N^+=4$ series which are coupled by the interaction mechanism of type (i-b).

As a final example, the strong feature associated with the transition to the 56f(2) state can be seen to ionize over two ranges of fields. The first of these encompasses fields from 30 to 50~V/cm, while the second ranges from 60 to 120~V/cm. This suggests that the long-lived states excited at this resonance fall within two distinct Rydberg series. These two sets of ionization fields correspond to those of hydrogenic states with $n=56$ and $N^+=2$, and $n=69$ and $N^{+} = 0$ (rectangle labelled C). The $N^+=2$ states are populated following $\ell$-mixing in residual uncanceled stray electric fields. The channel interaction mechanism that leads to the population of the $N^+=0$ states is that denoted type (i-b) in Section~\ref{sec:interactions}. As above, additional $\ell$-mixing by weak uncanceled electric fields results in the population of the long-lived hydrogenic states.

\begin{figure}
\includegraphics[width=0.49\textwidth]{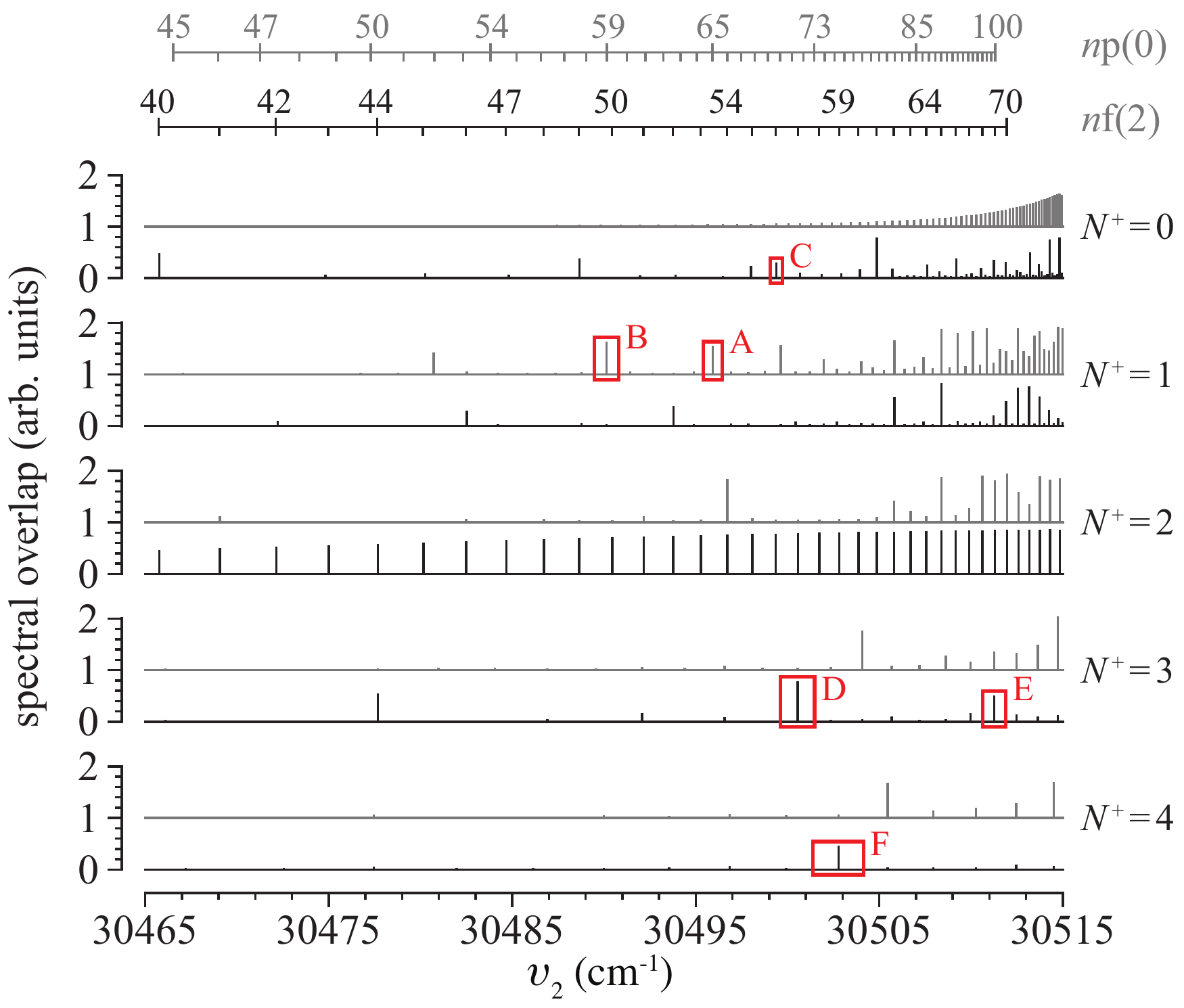}
\caption{\label{fig:stark_overlap} The spectral overlap between $n$p(0) [grey], and $n$f(2) [black] Rydberg states in NO with the hydrogenic states with $N^+=0$, 1, 2, 3 and~4 (see text for details). For each value of $N^+$ the $n$p(0) overlap is vertically offset for clarity of presentation.}
\end{figure}


\subsection{Numerical calculations}\label{simulation}

The complete interpretation of the experimental data in Figure~\ref{fig:wide_n_full} requires the combination of multichannel quantum defect theory~\cite{gauyacq90a,hiyama02a,goodgame_stark_2002} or matrix diagonalization methods~\cite{vrakking96a,bixon96a,goodgame_stark_2002,patel_observation_2007,seiler11b}, and a numerical treatment of the time-evolution of the excited state population in the stray electric fields, and 300~K blackbody radiation field in the apparatus~\cite{selier11a,seiler16a}. However, to highlight the primary role played by near degeneracies between $n$p(0) and $n$f(2) Rydberg states, which are accessible by electric dipole transitions from the A\,$^2\Sigma^+$ intermediate state, and high-$\ell$ hydrogenic states in the work reported here, we consider a simple numerical model in which seven Rydberg series are included. These are the $n$p(0) and $n$f(2) series with Hund's case (d) quantum defects of $\delta_{\mathrm{p}(0)}=0.7$ and $\delta_{\mathrm{f}(2)}=0.01$, respectively, and the series of long-lived hydrogenic Rydberg states converging to the $N^+=0$, 1, 2, 3 and 4 ionization thresholds and for which $\delta_{\ell(N^+)}=0$. Excitation of the long-lived hydrogenic states depends on their wavenumber proximity to states in the $n$p(0) and $n$f(2) series, and the intramolecular interactions discussed in Section~\ref{sec:interactions}. 

The spectral overlap between the $n$p(0) and $n$f(2) states and each series of hydrogenic states is estimated from the overlap of Lorentzian functions [0.018~cm$^{-1}$ full-width-at-half-maximum (FWHM) which was found to represent a crude approximation to the average wavenumber range of the intramolecular coupling between near degenerate p/f and high-$\ell$ hydrogenic states across the spectral region of interest] centred at the calculated wavenumbers of transitions to each series. This overlap is displayed in Figure~\ref{fig:stark_overlap}. The transition wavenumbers associated with the $n$p(0) and $n$f(2) states are indicated at the top of the figure. Beneath this are five sets of stick spectra. In the upper pair of these, the wavenumber associated with each spectral feature corresponds to the transition wavenumber to the long-lived hydrogenic Rydberg states with $N^+=0$. The intensity of each feature is the overlap integral of a set of Lorentzian functions, representing transitions to these hydrogenic $N^+=0$ states, with a second set of Lorentzian functions representing the transitions to the $n$p(0) (upper grey spectrum) and $n$f(2) (lower black spectrum) states. The spectral overlap of the long-lived hydrogenic states with the $n$p(0) states gradually increases as $n$ increases. On the other hand, the spectral overlap of the hydrogenic $N^+=0$ states with the $n$f(2) series, induced by the combination of the interaction mechanism denoted type (i-b) in Section~\ref{sec:interactions} and weak stray electric fields, does not change smoothly with $n$. Instead, in some particular cases this mixing is important but for most it is not. Many of the states for which mixing occurs can be identified in the experimental data. For example, the state highlighted by the rectangle labelled C in Figure~\ref{fig:stark_overlap} is also seen in Figure~\ref{fig:wide_n_full}.

A similar set of stick spectra are displayed in Figure~\ref{fig:stark_overlap} for the series of long-lived hydrogenic Rydberg states with $N^+=1$. In this case the couplings to the $n$p(0) and $n$f(2) series are induced by a combination of the type (i-a) interaction discussed in Section~\ref{sec:interactions} and weak stray electric fields. The spectral overlap does not vary smoothly with $n$ but instead exhibits pseudo-random occurrences of degeneracy. The features enclosed by the rectangles labelled A and B in the upper of these two spectra are also seen and assigned directly in the experimental data in Figure~\ref{fig:wide_n_full}.

The $N^+=2$ spectra in Figure~\ref{fig:stark_overlap} exhibit strong and consistent overlap with the $n$f(2) states. The mixing between these series is induced as a result of weak stray electric fields. The less significant overlap of the hydrogenic $N^+=2$ states with the $n$p(0) series is a result of the combination of the type (i-b) intramolecular interaction mechanism and further effects of weak electric fields. 

Finally, the intramolecular interaction mechanisms discussed in Section~\ref{sec:interactions} do not result in coupling between the $n$p(0) series and the series of long-lived hydrogenic Rydberg states with $N^+ = 3$ or~4. However, the type (i-a) interaction combined with effects of weak electric fields does lead to couplings of these series with the $n$f(2) series. In this case wavenumber coincidences between the series are rare but three cases are particularly noticeable. These (indicated by the rectangles labelled D, E and F) correspond directly to features assigned in the experimental data in Figure~\ref{fig:wide_n_full}.

\begin{figure}
\includegraphics[width=0.47\textwidth]{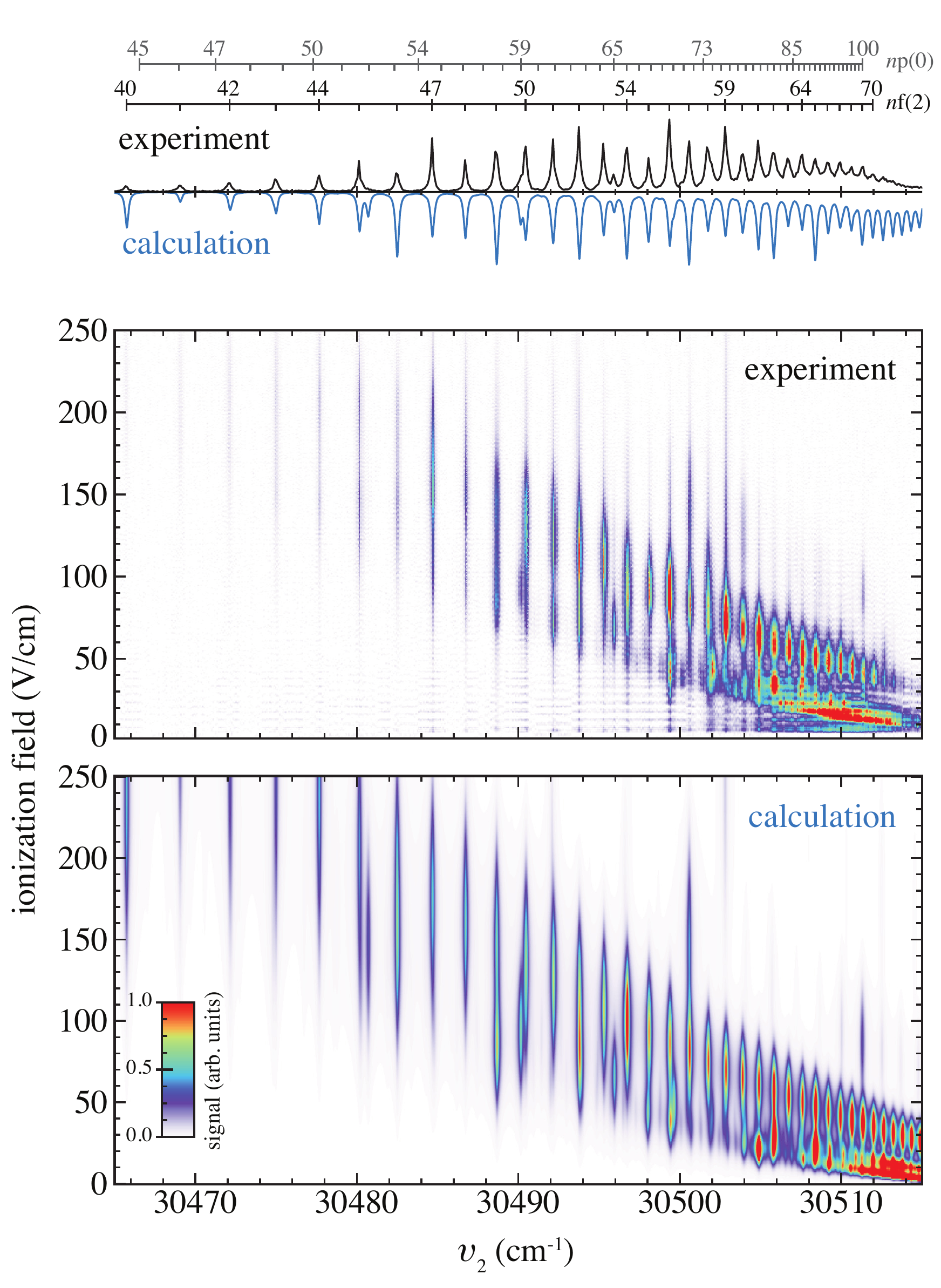}
\caption{\label{fig:wide_n_simulation} Experimental and calculated delayed electric field ionization signal of long-lived Rydberg NO molecules excited in nominally zero electric field as indicated. Spectra representing the total integrated signal at each laser wavenumber are displayed in the top panel along with the expected values of $\upsilon_2$ for the transitions to the $n$p(0) and $n$f(2) Rydberg series.}
\end{figure}

To make a quantitative comparison between experimental data and the intramolecular interaction mechanisms considered in the calculations in Figure~\ref{fig:stark_overlap} the absorption, and electric field ionization spectra were simulated. The electric field ionization spectra, displayed in Figure~\ref{fig:wide_n_simulation}, were obtained by first determining the range of ionization electric fields of the hydrogenic Rydberg-Stark states with each value of $n$. The amplitude of a Gaussian function encompassing this distribution of fields was then scaled according to the $n^{-3}$ dependence of the spectral intensity of the transitions to the $n$p(0) and  $n$f(2) states, the $n^3$ dependence of their lifetimes, and the spectral overlap of these states with the long-lived hydrogenic states in Figure~\ref{fig:stark_overlap}. The results were then convoluted with a Lorentzian function (0.12~cm$^{-1}$ FWHM) representing the spectral profile of the laser in the experiments. The absorption spectrum was determined by integrating the resulting electric field ionization signal over fields ranging from 0 to 250~V/cm.

The two-dimensional spectral maps used to identify the ionization fields and distinct Rydberg series in the experiments were calculated by combining the spectral intensities of the lower inverted spectrum at the top of Figure~\ref{fig:wide_n_simulation} with Gaussian distributions of ionization electric fields ranging from $F_{\mathrm{ion}}^{\mathrm{diabatic}^-}$ to $F_{\mathrm{ion}}^{\mathrm{diabatic}^+}$ for each set of hydrogenic states. Almost all of the main features observed in the experimental data are accurately reproduced in these calculations giving confidence in the interpretation of the experimental results and the mechanisms by which the long-lived excited states observed in the experiments were populated. 

\section{Discussion \& Conclusions}\label{sec:conc}

The high Rydberg states of NO studied here have lifetimes in excess of 10~$\mu$s. These long lifetimes and the electric field ionization data recorded in the experiments indicate that the states populated possess predominantly high-$\ell$, high-$|m_{\ell}|$ character. This hydrogenic character is attained through a combination of the intramolecular interaction mechanisms specific to the Rydberg states of NO, and effects of weak uncanceled static or time-dependent stray electric fields in the experimental apparatus. These fields arise from low-frequency electrical laboratory noise, and the presence and motion of ions in the photoexcitation region. From the amplitude of the measured NO$^+$ ion signal in the experiments, the fraction of initially excited molecules (number density $\sim10^7$~cm$^{-3}$) in these hydrogenic states is estimated to be $\sim10^{-2}$.

The long lifetimes and hydrogenic character of the excited states detected by delayed electric field ionization indicates that they are well suited to Rydberg-Stark deceleration and trapping experiments. At each wavenumber associated with resonant excitation, states with a range of static electric dipole moments are produced. These electric dipole moments can be as large as $(3/2)\,n^2\,ea_\mathrm{NO}$, where $a_\mathrm{NO}$ is the Bohr radius corrected for the reduced mass of NO, and, for example, have values approaching 12\,000~D for $n=56$. Using chip-based transmission-line Rydberg-Stark decelerators and electrostatic traps~\cite{lancuba_electrostatic_2016} it is expected that the molecules in states with electric dipole moments of this magnitude can be decelerated from 800~m/s to rest in the laboratory frame of reference within $\sim100~\mu$s. Electrostatic trapping experiments of this kind will allow the lifetimes of these Rydberg states to be determined through measurements of trap decay. Such measurements will provide new insight into slow decay processes of highly excited Rydberg states of small molecules. The implementation of the methods of Rydberg-Stark deceleration will also allow new merged-beam studies of ion-molecule reactions involving the NO$^+$ cation at low temperature.

\begin{acknowledgments}
We thank John Dumper and Rafid Jawad (UCL) for technical assistance, and Prof. Fr\'ed\'eric Merkt (ETH Zurich) for valuable discussions. This work was supported by the European Research Council (ERC) under the European Union's Horizon 2020 research and innovation program (grant agreement No. 683341).
\end{acknowledgments}

\bibliography{lib}

\end{document}